\documentclass[
    ,final            
  ]
  {aipproc}

\layoutstyle{6x9}

\begin{document}

\title{Point Processes Modeling of Time Series Exhibiting Power-Law Statistics}


\author{B. Kaulakys}{
  address={Institute of Theoretical Physics and Astronomy of Vilnius University, A. Gostauto 12, LT-01108 Vilnius, Lithuania}
}

\author{M. Alaburda}{
  address={Institute of Theoretical Physics and Astronomy of Vilnius University, A. Gostauto 12, LT-01108 Vilnius, Lithuania}
}

\author{V. Gontis}{
  address={Institute of Theoretical Physics and Astronomy of Vilnius University, A. Gostauto 12, LT-01108 Vilnius, Lithuania}
}

\begin{abstract}We consider stochastic point processes generating time series exhibiting power laws of spectrum and distribution density (\emph{Phys. Rev. E} \textbf{71}, 051105 (2005)) and apply them for modeling the trading activity in the financial markets and for the frequencies of word occurrences in the language. 
\end{abstract}

\maketitle


\section{Introduction}

Recently, we proposed \cite{Kaulakys1998} and generalized \cite{Kaulakys2005} the stochastic point process models generating a variety of monofractal and multifractal time series exhibiting power laws of the spectrum $S(f)$ and of the distribution $P(I)$ of the signal intensity and applied them for the analysis of the financial systems \cite{Gontis2004}. These models can generate $1/f$ noise with a very large Hooge parameter. They may be used as the theoretical framework for understanding huge fluctuations (see, e.g., \cite{Kruppa2006}) and as well for description of a large variety of observable statistics, i.e., jointly power spectral density (PSD), $S(f)\sim 1/f^{\beta}$, signal probability distribution function (PDF), $P(I)\sim 1/I^{\lambda}$, with different slopes, different distributions, $P(\tau)$, of the interevent time $\tau$ and different multifractality. 

Here we will present the extensions and generalizations of the point process models for the Poissonian-like processes with slowly diffusing mean interevent time \cite{Gontis2006}. We will adjust the parameters of the generalized model to the empirical data of the trading activity in the financial markets \cite{Gontis2007} and to the frequencies of the word occurrences in the language, reproducing the PDF and PSD. 

\section{The Model }

We investigate stochastic time series as a sequence of events which occur at discrete times ${t_{1},t_{2},...t_{k},...}$ and can be considered as identical point events. Such point process equivalently is defined by the set of stochastic interevent times $\tau_{k}=t_{k+1}-t_{k}$. Let us consider the flow of events as the Poissonian-like process driven by the multiplicative stochastic equation. We define the stochastic rate $n=1/\tau$ of event flow by continuous stochastic differential equation
\begin{equation}
\mathrm{d}\tau=\left[\gamma-\frac{m}{2}\sigma^2\left(\frac{\tau}{\tau_{0}}\right)^m\right]\tau^{2\mu-2}\mathrm{d}t+\sigma\tau^{\mu-1/2}\mathrm{d}W,
\label{eq:taustoch2}
\end{equation}
where $W$ is a standard Wiener process, $\sigma$ denotes the standard deviation of the white noise, $\gamma\ll 1$ is a coefficient of the nonlinear damping and $\mu$ defines the power of noise multiplicativity. The diffusion of $\tau$ is restricted from the side of high values by an additional term $-\frac{m}{2}\sigma^2\left(\frac{\tau}{\tau_{0}}\right)^m\tau^{2\mu-2}$, which produces the exponential diffusion reversion. $m$ and $\tau_{0}$ are the power and value of the diffusion reversion, respectively. The associated Fokker-Plank equation with the zero flow gives the simple stationary PDF 
\begin{equation} 
P(\tau)\sim\tau^{\alpha+1}\exp\left[-\left(\frac{\tau}{\tau_{0}}\right)^m\right]\label{eq:taudistrib} 
\end{equation} 
with $\alpha=2(\gamma_{\sigma}-\mu)$ and $\gamma_{\sigma}=\gamma/\sigma^2$. Eq. (\ref{eq:taustoch2}) describes continuous stochastic variable $\tau$, defines rate $n=1/\tau$ with stationary distribution and PSD $S(f)$ \cite{Gontis2006,Gontis2007}, 
\begin{equation}
P(n)\sim\frac{1}{n^{\lambda}}\exp\left\{-\left(\frac{n_{\mathrm{0}}}{n}\right)^m
\right\},\quad\lambda=1+\alpha,\label{eq:ndistr}
\end{equation}
\begin{equation}
S(f)\sim\frac{1}{f^{\beta}},\quad\beta=1+\frac{\alpha}{3-2\mu}.
\label{eq:nspekt}
\end{equation}
Here we define the fractal point process driven by the stochastic differential equation (\ref{eq:taustoch2}), i.e., we   assume $\tau(t)$ as slowly diffusing mean interevent time of the Poissonian-like process with the stochastic rate $n$. 
Within this assumption the conditional probability of interevent time $\tau_{\mathrm{p}}$ in the Poissonian-like process with the stochastic rate $1/\tau$ is 
\begin{equation}
\varphi(\tau_{\mathrm{p}}|\tau)=\frac{1}{\tau}\exp\left[-\frac{\tau_{\mathrm{p}}}{\tau}\right].\label{eq:taupoisson}
\end{equation}
Then the long time distribution $P_k(\tau_{\mathrm{p}})$ of interevent time $\tau_{\mathrm{p}}$ in $k$-space \cite{Kaulakys2005} has the integral form 
\begin{equation}
P_k(\tau_{\mathrm{p}})=C\int_{0}^{\infty}\exp\left[-\frac{\tau_{\mathrm{p}}}{\tau}\right]\tau^{\alpha-1}\exp\left[-\left(\frac{\tau}{\tau_{0}}\right)^m\right]\mathrm{d}
\tau,\label{eq:taupdistrib}
\end{equation}
with $C$ defined from the normalization, $\int_{0}^{\infty}P_k(\tau_{\mathrm{p}})\mathrm{d}
\tau_{\mathrm{p}}=1$. The distributions of interevent time $\tau_{\mathrm{p}}$ have their explicit forms for the integer values of power $m$. For $m=1$ and for $m>1$ they are expressed by the modified Bessel function \cite{Gontis2006,Gontis2007} and in terms of the hypergeometric functions, respectively.

\section{Discussion }

\begin{figure}
  \includegraphics[width=.45\textwidth]{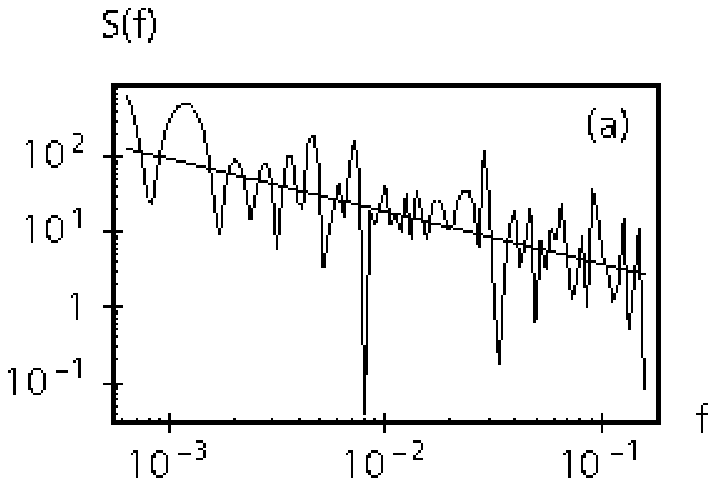}
  \includegraphics[width=.45\textwidth]{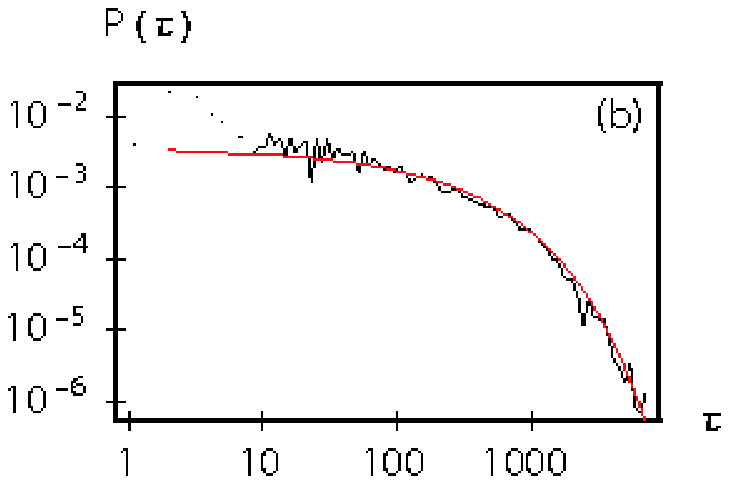}
  \caption{(a) The power spectral density $S(f)$ of the point process for word "eye" occurrences in the novels of Jack London. The straight line approximates the power-law with the exponent $\beta=0.7$. (b) The interevent interval $\tau$ distribution of the word "eye" occurrences calculated from the histogram in the same novels. The smooth line represents the integral formula (\ref{eq:taupdistrib}) with $m=2$, $\tau_{0}=800$ and $\alpha=-0.3$.}
\label{fig:1}
\end{figure}
We consider two applications of the proposed model. The frequencies of the word occurrences in the language depend on the content and diffuse in the text. We consider the flow of words in the text as the steps of discrete events, i.e., one word is the unit of the interval. Then the number of other words in between of the two successive occurrences of the same noun measures the interevent interval $\tau_{\mathrm{p}}$ of the point process defined for the sequence of selected noun. One can easily calculate the sequence of all selected word occurrence intervals $\tau_{\mathrm{p}}$ and so define the realization of the point process. Here we demonstrate the statistics of the word "eye" in the selected novels of Jack London, over 1.2 mln. words totally. First of all, we demonstrate that the point process, defined in such a way, has long memory as the exponent of PSD $\beta=0.7$ [Fig. 1 (a)]. With the assumption of pure multiplicative process with $\mu=1$ from the relation $\beta=2\gamma_{\sigma}-1=1+\alpha$ one defines the parameter $\alpha=-0.3$ related with the probability distribution functions. The histogram of $\tau_{\mathrm{p}}$ distribution coincides with theoretical PDF defined by its integral form (\ref{eq:taupdistrib}) when $m=2$ [Fig. 1 (b)]. The exponent $\lambda$ of the power-law distribution for number of the word "eye" occurrences in the 1000 words length pieces of text is $\lambda=4.4$. As we will see later, the presented example of the word statistics resembles the statistical properties of trading activity in the financial markets.
\begin{figure}
  \begin{minipage}{\columnwidth}
  \includegraphics[width=.45\textwidth]{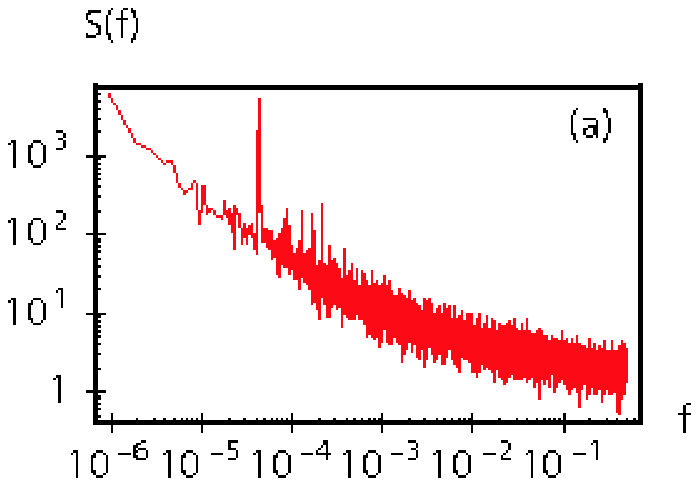}
  \includegraphics[width=.45\textwidth]{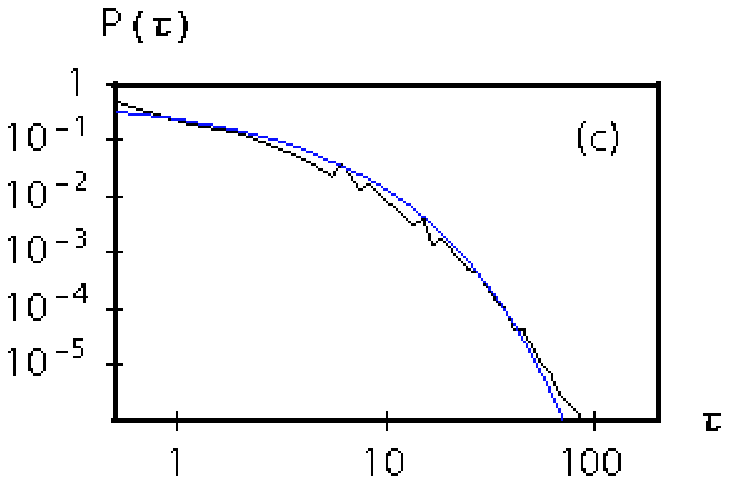}\newline
  \includegraphics[width=.45\textwidth]{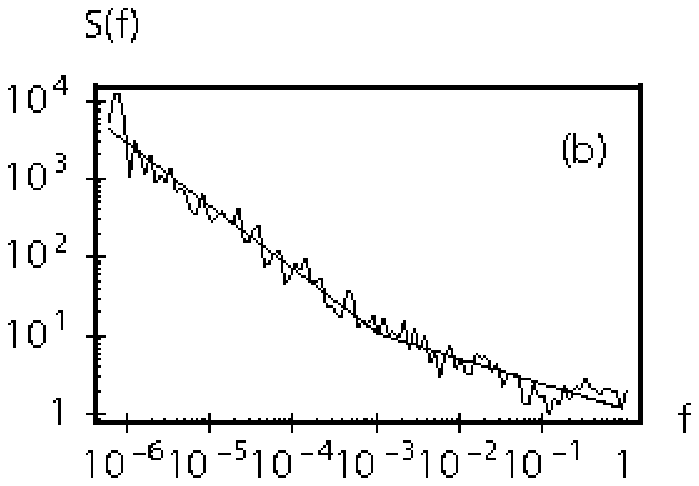}
  \includegraphics[width=.45\textwidth]{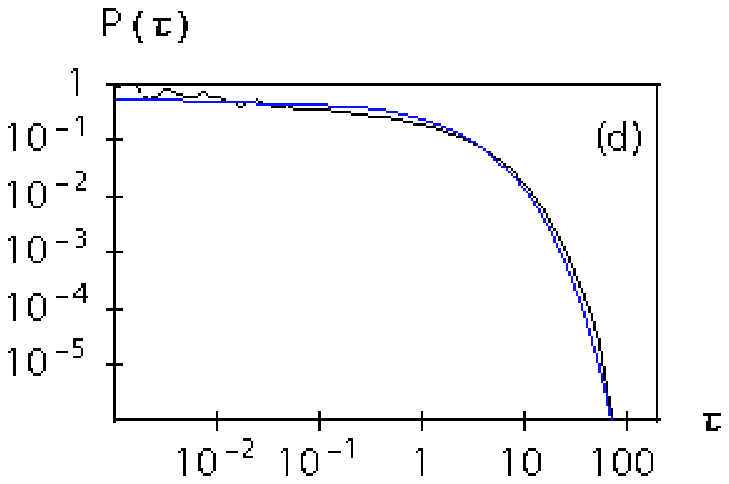}
  \end{minipage}
  \caption{(a) The power spectral density $S(f)$ of the stock CVX trade sequence traded on the NYSE. (b) Power spectral density of the Poissonian-like process driven by Eq. (\ref{eq:taucontinuous}) with the parameters $\tau_{0}=4$, $\sigma=0.023$, $\gamma=0.00034$, and $\epsilon=0.06$. Straight lines approximate power-law spectrum with exponents $\beta_{1}=0.8$ and $\beta_{2}=0.3$. (c) The empirical distribution of $\tau_{p}$ calculated from the histogram of CVX trades on NYSE. The smooth line represents the integral formula (\ref{eq:taupdistrib}) for $m=2$, $\tau_{0}=4$ and $\alpha=-0.3$. (d) The model distribution of $\tau_{p}$ calculated with same parameters as in (b), the smooth line is the same as in (c).} 
\label{fig:2}
\end{figure}

In the case of the financial market we consider every trade of the selected stock as a point event, i.e. the sequence of all trades for the stock composes the stochastic point process, described by the set of time intervals between the successive trades. The power spectral density of the trade sequence serves as a measure of the long range memory property of trading activity. An example of the spectrum for the stock CVX trade sequence traded in the period of two years on NYSE, Fig.2 (a), reveals the structure of the power spectral density in a wide range of frequencies and shows that the real markets exhibit two power laws with the exponents $\beta_{1}=0.8$ and $\beta_{2}=0.3$. In our recent works \cite{Gontis2006,Gontis2007} we have proposed the model adjustment, introducing a new form of the modulating stochastic differential equation instead of Eq. (\ref{eq:taustoch2}), 
\begin{equation}
\mathrm{d}\tau=\left[\gamma-\frac{m}{2}\sigma^2\tau^m\right]\frac{1}{(\epsilon+\tau)^2}\mathrm{d}t
+\sigma\frac{\sqrt{\tau}}{\epsilon+\tau}\mathrm{d}W,
\label{eq:taucontinuous}
\end{equation}
where a new parameter $\epsilon$ defines the crossover between two areas of $\tau$ diffusion with assumption $\tau_{0}=1$. The solution of Eq. (\ref{eq:taucontinuous}) has to be scaled by $\tau_{0}$ for other values of $\tau_{0}$. The Poisonian-like point process modulated by Eq. (\ref{eq:taucontinuous}) reproduces PSD, Fig.2 (a), of the empirical trade sequence in detail, including two exponents and the crossover point, Fig.2 (b). The proposed model with the same parameters reproduces the empirical PDF of $\tau_{p}$, Fig.2 (c) and (d), very well. 
Moreover, the model describes the distribution of the empirical trading activity, i.e., the number of transactions per selected time window with the power-law exponent $\lambda=4.4$.

\begin{theacknowledgments}
We acknowledge the support by the Agency for International Science and Technology Development Programs in Lithuania and EU COST Action P10 ``Physics of Risk''.
\end{theacknowledgments}



\bibliographystyle{aipproc}   


\IfFileExists{\jobname.bbl}{}
 {\typeout{}
  \typeout{******************************************}
  \typeout{** Please run "bibtex \jobname" to optain}
  \typeout{** the bibliography and then re-run LaTeX}
  \typeout{** twice to fix the references!}
  \typeout{******************************************}
  \typeout{}
 }



\end{document}